\documentstyle[sprocl]{article}
\def\beq{\begin{equation}}
\def\eeq{\end{equation}}
\def\beqn{\begin{eqnarray}}
\def\eeqn{\end{eqnarray}}
\begin{document}
\title{
DPF '96: THE TRIUMPH OF THE STANDARD MODEL}
\author{ S. DAWSON}
\address{Physics Department\\
Brookhaven National Laboratory
\\
Upton, N.Y.  11973}
\maketitle\abstracts{
I  summarize some of the  highlights of the 1996 DPF meeting, 
paying   
particular attention to new measurements of the $W$, $Z$,
and top quark masses.
Precision electroweak measurements from LEP are discussed with emphasis 
on recent  measurements of $R_b$
and values of the coupling constants $\alpha(M_Z^2)$ and $\alpha_s(M_Z^2)$
are presented.
Taken as a whole, the data are in spectacular agreement with the
predictions of the Standard Model.  
}
\section{Introduction} 
This meeting saw many beautiful experimental
 results presented, the overwhelming majority of
which  support
the correctness of our basic understanding  of particle physics.
Many of the puzzles
and data which did not fit
into our picture  from last year's  conferences have 
become less compelling, 
leaving a wealth of data forming a consistent picture.  The first
observations of $W$ pairs from LEP were
presented, along with new measurements of the $W$, $Z$, and top 
quark masses.  The errors on
all of these masses are significantly reduced
from previous values.  Numerous  electroweak precision measurements
were presented, along with new measurements of
$\alpha(M_Z^2)$ and  $\alpha_s(M_Z^2)$.
  
In this note, I  give a (very
subjective!)  lightning  review of  some of the highlights
of the meeting.  Unfortunately, there are many exciting 
and important results which  I will not be able to cover.  
This has been truly a productive year for particle physics.  
\section{Precision Measurements of Masses}
\subsection{The $Z$ Mass}

The mass of the $Z$ boson is usually taken as an
input parameter in studies of electroweak physics.  
At the $Z$ resonance, 
the error on the $Z$ mass is directly related to the
precision with which the beam energy is measured.
Previous measurements have taken into account the
phases of the moon and the weight of Lake Geneva
on the ring.  The latest measurement incorporates
 the time schedules 
 of the TGV trains
(which generate vagabond currents  in the beam)
 and leads to a measurement
with errors\cite{tc} 
\beqn
\Delta M_Z&=&\pm 1.5~MeV  \\ \nonumber
\Delta \Gamma_Z&=& \pm 1.7~MeV~\quad .   
\eeqn
These errors yield  a new 
combined LEP result from the $Z$ lineshape,~\cite{tc} 
\beqn 
M_Z&=&  91.1863\pm .0020~GeV 
\\ 
\nonumber 
\Gamma_Z&=& 2.4946\pm .0027~GeV\quad .  
\qquad\qquad {\rm LEP}  
\eeqn 
The $Z$ mass is down $4~MeV$ from the previous measurement.  
This shift is due almost entirely to understanding the effects
of the trains! 
\subsection{The $W$ Mass}
 
The LEP experiments have  presented
preliminary  measurements of the $W$ pair
production cross section.  $W^+W^-$ pairs have been observed in the
$q {\overline q} q {\overline q}$,
$q {\overline q} l \nu$, and $ l \nu l \nu$ decays modes, with the
number 
of $W$ pairs increasing daily.  Because of the sharp threshold behaviour
of the production cross section, $\sqrt{s}
\sim 161~GeV$ is the optimal
energy at which to measure the $W$ mass and the $M_W$ dependance of
the cross section at this point is relatively insensitive to new physics
effects.  The combined result from the $4$ LEP experiments at 
$\sqrt{s}=161.3 \pm .2~GeV$ is,~\cite{gb}  
\beq
\sigma(e^+e^-\rightarrow W^+W^-)=3.6\pm .7~pb
\quad . 
\qquad\qquad {\rm LEP} 
\eeq
Assuming the validity of the Standard Model, this gives a
new measurement of the $W$ mass,\cite{gb} 
\beq 
M_W=80.4\pm.3\pm.1~GeV
\quad .  \qquad\qquad {\rm LEP}
\eeq
Since the error is dominated by statistics, it
should be reduced considerably with further running.
 The data presented correspond to $3~ pb^{-1}$ per experiment.  

$W^+W^-$ pair production at LEP will also be used to measure 
deviations of the $W^+W^-Z$ and 
$W^+W^-\gamma$ couplings from their
Standard Model values and OPAL presented preliminary
limits on these couplings 
 as a ``proof of principle".\cite{gb}     
 These limits are not yet competitive with those obtained
at the Tevatron.   
  
The D0 collaboration  presented a new measurement of the $W$ mass
from the
transverse mass spectrum of $W\rightarrow e \nu$, \cite{ak}
\beq
M_W=80.37\pm .15~ GeV\quad .  \qquad \qquad {\rm D0} 
\eeq 
This error is considerably smaller than previous CDF and D0 $W$ mass
measurements. 
These results contribute  to 
a new world average,\cite{md} 
\beq
M_W=80.356\pm .125~GeV\quad . \qquad \qquad {\rm WORLD}
\eeq 
\subsection{The Top Quark Mass}
  
The top quark has moved from being a newly discovered particle to
 a
mature particle whose properties  can be  studied in
detail.  CDF and D0 each have more than
$100~pb^{-1}$ of data which means that about $500~t
{\overline t}$ pairs have been produced
in each experiment.  Together, the experiments have
identified
 around $13$ di-lepton, $70$ lepton
 plus jets, and $60$ purely hadronic
 top events and the top
quark  cross section and mass have been measured in many
channels.  
The cross sections and masses obtained from the
various channels are in good agreement and 
  the combined results from CDF and D0 
at $\sqrt{s}=1.8~TeV$ are, \cite{bw,top} 
\beqn
\sigma_{t {\overline t}}&=& 6.4_{-1.2}
^{+1.3}~pb \nonumber \\
M_T&=& 175 \pm 6~GeV\quad . \qquad \qquad 
{\rm CDF,D0} 
\eeqn
The error on $M_T$ of $\pm 6~GeV$ is a factor of $2$ smaller than that
reported in February, 1995 due both 
to greater statistics and to improved analysis techniques. 
 The dominant source of error
remains the jet energy correction.

There  has been 
considerable theoretical effort devoted to computing the
top quark cross section in QCD beyond the leading order.  In
order to sum the soft gluon effects, (which are numerically
important), the non-perturbative regime must be confronted,
leading to some differences between the various calculations.\cite{eb} 
The theoretical cross section is slightly higher than the
experimental value, but is in reasonable agreement.  

The direct measurement of $M_T$ can be compared with the
indirect 
result inferred from precision electroweak
measurements at LEP and SLD,\cite{md,pl} 
\beq
M_T=179\pm 7^{+16}_{-19}~GeV\quad .\qquad\qquad {\rm INDIRECT} 
\eeq
(The second error results from varying the Higgs mass between 
$60$ and $1000$ GeV with the central value taken as $300~GeV$.)
This is truly an impressive agreement between the direct
and indirect measurements!  

Measurements of the top quark properties can be used to probe
new physics.  For example, by measuring the branching ratio of
$t\rightarrow W b$ (and assuming $3$ generations
of quarks  plus unitarity),
the $t-b$ element of the 
Kobayashi-Maskawa matrix can be measured,\cite{bw,ts} 
\beq
\mid V_{tb}\mid =.97\pm.15\pm.07\quad .\qquad\qquad
{\rm CDF}
\eeq	 
\section{Precision Electroweak Measurements}

There were many  results from precision 
electroweak measurements presented at this meeting, most of
which are in spectacular agreement with the predictions
of the Standard Model.  (See the talks by 
P.~Langacker~\cite{pl} and
M.~Demarteau~\cite{md} for tables of electroweak measurements
and the comparisons with Standard Model predictions).
Here, I will discuss two of those measurements,
\beqn
R_b&\equiv &{\Gamma(Z\rightarrow b {\overline b})
\over \Gamma(
Z\rightarrow {\rm hadrons})} 
\nonumber \\
A_b&\equiv &  
{2 g_V^bg_A^b\over
[(g_V^b)^2+(g_A^b)^2]}
\qquad . 
\eeqn 
Both of these measurements differ from the Standard Model
predictions 
 and are particularly interesting theoretically
since they involve the couplings of the third generation quarks. 
In many non-standard
models, the effects of new physics would first show
up in the couplings of gauge bosons to the $b$ and $t$ quarks. 

A year ago, the value of $R_b$ was about $3\sigma$
above the Standard Model prediction.  At this meeting new results
were presented by the SLC collaboration and by the $4$ LEP
experiments.  
Numerous improvements in the analyses have been 
made, including measuring many of the charm decay rates directly
instead of inputting values from other experiments.
The ALEPH and SLD experiments
have employed   a new analysis  technique
utilizing a lifetime and mass tag.  This technique allows them
to obtain $b$ quark samples which are $\sim 97\%$ pure,
while maintaining relatively high efficiencies.
  This purity is considerably larger than
that obtained in previous studies of $R_b$.  
The new ALEPH~\cite{ab}
 and SLD~\cite{ew} results are right on the nose of
the Standard Model prediction,
\beq
R_b=\left\{
\begin{array}{ll}
.21582\pm.00087{\rm(stat)}
&
{\rm ALEPH} 
 \\
.2149\pm.0033 {\rm (stat)}\pm.0021 {\rm(syst)}
\pm.00071 {\rm (R_c)}  &
{\rm SLD} 
 \\
.2156\pm .0002
\qquad . 
&
{\rm SM}
\end{array}
\right. 
\eeq 
(The theory error results from varying $M_H$).\cite{pl}  
Incorporating all measurements leads to a new world
 average, \cite{md,pl}
\beq
R_b=.2178\pm .0011
\quad ,  
\qquad\qquad {\rm WORLD} 
\eeq
which is $1.8\sigma$ above the Standard Model.  
Advocates of supersymmetric models remind us that it is
difficult to obtain effects larger than $2\sigma$ in these
models, so the
possibility that $R_b$ may indicate new physics remains, although
the case for it has certainly been  weakened.  
 (The value of $R_c$ is now within $1\sigma$ of the Standard 
Model prediction.)

The only electroweak precision measurement which 
is in serious disagreement with the Standard Model
prediction  is $A_b$,
which is sensitive to the axial vector coupling of the 
$b$ quark.  The new SLD result
obtained using a lepton sample,\cite{gm} 
\beq
A_b=.882\pm.068 {\rm (stat)}\pm
.047 {\rm (syst)}\qquad \qquad {\rm SLD}  
\eeq
leads to a revised  world average,
\beq
A_b=.867\pm .022\quad,  \qquad\qquad {\rm WORLD}
\eeq
about $3\sigma$ below the Standard Model prediction of $A_b=.935$.  
There are,however,
 assumptions involved in comparing the SLD and LEP
numbers which may help resolve this
discrepancy.\cite{jh}  
 
The LEP and SLD electroweak precision measurements can also be
used to infer a preferred value for the Higgs mass,
(including also the direct measurement of $M_T$ as an input),
\cite{md} 
\beq
M_H=149^{+148}_{-82~GeV}
\quad . \qquad {\rm INDIRECT}  
\eeq
This limit is driven by $R_b$ and $A_{LR}$.   
Since the observables depend only logarithmically
on $M_H$, there are large errors, but  it is interesting
that a relatively light value of $M_H$ seems to
be preferred.    
Such a light Higgs boson mass is
predicted    in supersymmetric
theories.  
 
The electromagnetic coupling constant can also be
extracted from electroweak precision measurements,
\beq
{1\over \alpha_{EM}(M_Z^2)}=128.894\pm .090
\qquad . 
\eeq 
This leads to an error of $\delta\sin^2\theta_W=.00023$,
which is roughly the same size as the experimental error.  
This emphasizes the need for a more precise 
measurement of $\alpha_{EM}$. 
\section{QCD and Measurements of $\alpha_s$}  

At the summer meetings a year ago, it seemed that the values  of
$\alpha_s(M_Z^2)$ as extracted from
lattice calculations and  low energy experiments were  smaller  
than the values extracted from measurements at the $Z$ pole.  This led to
numerous speculations of the possibilities for new physics to
cause this effect.  At this meeting the CCFR collaboration presented a new
measurement of $\alpha_s(M_Z^2)$
 obtained by fitting the $Q^2$ dependance of
the $\nu$ deep inelastic structure functions,
$F_2$ and $xF_3$, 
\cite{pss} 
\beq
\alpha_s(M_Z^2)=.119\pm.0015 
{\rm (stat)}\pm.0035 {\rm (syst)}
\pm .004 {\rm (scale)}
\qquad . 
\qquad {\rm CCFR}
\eeq
This value is higher than the
previous values of $\alpha_s(M_Z^2)$ extracted from
deep inelastic scattering experiments.  
We can compare  with the value extracted from the lineshape
at LEP~\cite{pl}
\beq
\alpha_s(M_Z^2)=.123\pm.004\qquad\qquad {\rm LEP}
\eeq
to see that there does not seem to be any systematic discrepancy
between the values of $\alpha_s(M_Z^2)$ measured at different energies.
 A   world average for $\alpha_s(M_Z^2)$ (not including
the new CCFR point) can be found, \cite{md} 
\beq
\alpha_s(M_Z^2)=.121\pm.003\pm.002\qquad . \qquad\qquad {\rm WORLD}
\eeq
Most of the extracted values of $\alpha_s(M_Z^2)$ are within
$1\sigma$ of this value.\cite{pl}

The inclusive jet cross sections measured at the Tevatron
continue to show an excess  of events
at high $E_T$ when compared with the theoretical
predictions.\cite{etjet}  When corrections are made 
for differences in the rapidity coverages, etc,
between the detectors,   
 the CDF and D0 data on inclusive
jet cross sections  are in agreement.\cite{ebg} 
The data can be partially explained by adjusting the gluon structure
function at large $x$,\cite{hl}
 although considerable theoretical work remains to
be done before this effect is completely understood. 

\section{$\nu$ Puzzles}

The deficit of solar neutrinos from the Homestake mine, Kamiokande,
SAGE, and GALLEX
experiments remains a puzzle,  as
it is not easily explained by adjustments to the
solar model.  These results could be
interpreted in terms of oscillations.\cite{hs}
 The LSND
collaboration presented positive
evidence for the oscillation ${\overline{
\nu_\mu}}\leftrightarrow {\overline{\nu_e}}$.~\cite{hk}
  They now have $22$ events
with an expected background of $4.6\pm.6$.  Their claim is
that the excess events are consistent with the
 oscillation hypothesis.
Hopefully, an upgraded KARMEN detector will be able to clarify the
LSND results.\cite{sm} 

\section{The $\tau$ lepton, $b$ and $c$ quarks}

This summary would not be complete without mentioning the
$\tau$, $b$ and $c$.  Although each of these
particles was discovered some years ago, interesting new
results on lifetimes, branching ratios, and mixing
angles continue to be reported.  See the reviews by
H.~Yamamoto~\cite{hy} and P.~Sphicas.\cite{ps}  

\section{New Physics}

There were many talks at this meeting
devoted to searches for physics beyond the Standard Model.
They can best be summarized by stating that 
there is no experimental evidence for such physics.
Many theorist's favorite candidate for 
physics beyond the Standard Model  is supersymmetry
and there were  a large number of  parallel talks with limits on the SUSY
spectrum, (see the reviews by W.~Merritt\cite{wm}
 and M.~Schmitt\cite{ms}).
In many cases, the limits are in the
interesting  $100-200~GeV$ range and seriously 
restrict models with supersymmetry at the electroweak scale. 

Considerable attention has been paid to a single CDF event
with  an $e^+e^-\gamma\gamma$ in the final state, along with
missing energy.  This event is particularly clean and lends
itself to various supersymmetric interpretations.  At this meeting,
however, the $E_T^{\rm miss}$ distribution in the $\gamma\gamma$
spectrum was  presented by the
CDF collaboration  and there is no additional evidence
(besides this one event) for unexplained physics
in this channel.\cite{dt} 

\section{Conclusions}  
  The theoretical predictions and experimental data discussed 
at this meeting form a coherent picture in which  the
predictions of the standard $SU(3)\times
SU(2)\times U(1)$ model have been validated many, many times.
We need to remind ourselves, however, that this is not the end
of particle physics and that
there are large areas in which we  continue to be almost
totally 
ignorant.  There remain many unanswered questions:
"How is the electroweak symmetry broken?", "Why are there three
generations of quarks and leptons?", "Why do the coupling constants
 and masses
have their measured values?" ....~The list goes on and on and our
questions can only be answered by future experiments.  
  
\section*{Acknowledgments} 
I am grateful to all the speakers who so generously
shared their transparencies and knowledge with me.  
This manuscript has been authored under contract number
DE-AC02-76CH00016 with the U.S. Department of Energy.
Accordingly, the U.S. Government retains a non-exclusive, royalty-free
license to publish or reproduce the published form of this contribution, or allow others to do so,
for U.S. Government purposes.

\end{document}